\documentclass[twocolumn,aps,prl,superscriptaddress,preprintnumbers,bibnotes]{revtex4-2}

\usepackage[T1]{fontenc}

\usepackage{natbib}
\usepackage{graphicx}
\usepackage{bm}
\usepackage[dvipsnames]{xcolor}
\usepackage{amsmath}
\usepackage{gensymb} 
\usepackage{physics}
\usepackage{tabularx}
\usepackage{ragged2e}
\usepackage{multirow}
\usepackage{bigdelim}
\usepackage{amssymb}
\usepackage[normalem]{ulem}
\usepackage{hyperref}
\hypersetup{colorlinks=true,linkcolor=Blue,filecolor=Blue,citecolor=Blue,urlcolor=Blue}

\newcommand\unit[2]{\ensuremath{#1\,\mathrm{{#2}}}}

\begin{document}

\title{Coherent excitation of a \textmu Hz scale optical magnetic quadrupole transition}

\author{V. Kl{\"u}sener}
\author{S. Pucher}
\affiliation{
  Max-Planck-Institut f{\"u}r Quantenoptik,
  85748 Garching, Germany}
\affiliation{
  Munich Center for Quantum Science and Technology,
  80799 M{\"u}nchen, Germany}
\author{D. Yankelev}
\thanks{Current address: Rafael Ltd., Haifa 3102102, Israel}
\affiliation{
  Max-Planck-Institut f{\"u}r Quantenoptik,
  85748 Garching, Germany}
\affiliation{
  Munich Center for Quantum Science and Technology,
  80799 M{\"u}nchen, Germany}
\author{J. Trautmann}
\author{F. Spriestersbach}
\affiliation{
  Max-Planck-Institut f{\"u}r Quantenoptik,
  85748 Garching, Germany}
\affiliation{
  Munich Center for Quantum Science and Technology,
  80799 M{\"u}nchen, Germany}
\author{D. Filin}
\author{S. G. Porsev}
\author{M. S. Safronova}
\affiliation{
  Department of Physics and Astronomy,
  University of Delaware,
  Newark, Delaware 19716, USA}
\author{I. Bloch}
\affiliation{
  Max-Planck-Institut f{\"u}r Quantenoptik,
  85748 Garching, Germany}
\affiliation{
  Fakult{\"a}t f{\"u}r Physik,
  Ludwig-Maximilians-Universit{\"a}t M{\"u}nchen,
  80799 M{\"u}nchen, Germany}
\affiliation{
  Munich Center for Quantum Science and Technology,
  80799 M{\"u}nchen, Germany}
\author{S. Blatt}
\email{sebastian.blatt@mpq.mpg.de}
\affiliation{
  Max-Planck-Institut f{\"u}r Quantenoptik,
  85748 Garching, Germany}
\affiliation{
  Fakult{\"a}t f{\"u}r Physik,
  Ludwig-Maximilians-Universit{\"a}t M{\"u}nchen,
  80799 M{\"u}nchen, Germany}
\affiliation{
  Munich Center for Quantum Science and Technology,
  80799 M{\"u}nchen, Germany}

\date{\today}

\begin{abstract}
We report on the coherent excitation of the ultranarrow $^{1}\mathrm{S}_0$--$^{3}\mathrm{P}_2$ magnetic quadrupole transition in $^{88}\mathrm{Sr}$. By confining atoms in a state insensitive optical lattice, we achieve excitation fractions of $97(1)\%$ and observe linewidths as narrow as \unit{58(1)}{Hz}. With Ramsey spectroscopy, we find coherence times of \unit{14(1)}{ms}, which can be extended to \unit{266(36)}{ms} using a spin-echo sequence. We determine the linewidth of the M2 transition to 24(7)\,\textmu Hz, confirming longstanding theoretical predictions. These results establish an additional clock transition in strontium and pave the way for applications of the metastable $^{3}\mathrm{P}_2$ state in quantum computing and quantum simulations.
\end{abstract}

\maketitle

The coherent excitation of ultra-narrow optical transitions between long-lived atomic states is fundamental for optical atomic clocks~\cite{ludlow15}, quantum information processing~\cite{madjarov20,lis23,haeffner08}, and quantum simulation~\cite{gross17,roos12}.

\begin{figure}
  \includegraphics{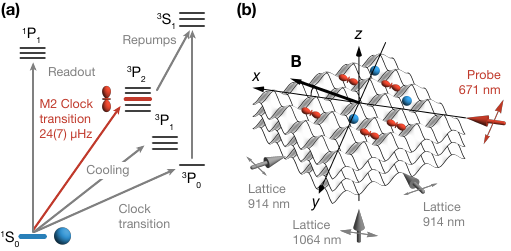}
  \caption{Level scheme of $^{88}\mathrm{Sr}$ and experimental setup. (a) Energy levels and transitions relevant for state preparation, spectroscopy and detection. (b) Atoms are confined in a three-dimensional optical lattice at lattice wavelengths of \unit{914}{nm} and \unit{1064}{nm} for the horizontal and vertical axes, respectively. The polarizations of the lattice beams are adjusted to angles $\beta_{m,914}$ and $\beta_{m,1064}$ relative to the quantization axis defined by the magnetic field $\mathbf{B}$ in the \textit{xz}-plane. The probe laser propagates with wave vector $\mathbf{k}$ along the \textit{x}-axis and is linearly polarized along \textit{y} perpendicular to the plane spanned by $\mathbf{k}$ and $\mathbf{B}$.}
  \label{fig:setup}
\end{figure}

The exceptional frequency resolution offered by such transitions provides the basis for the remarkable stability and precision of optical atomic clocks.
Work on divalent systems has so far almost exclusively focused on the $^{1}\mathrm{S}_0$--$^{3}\mathrm{P}_0$ clock transition due to its insensitivity to environmental perturbations.
As shown in Fig.~\ref{fig:setup}(a), a complementary clock transition with \textmu{}Hz-scale natural linewidth is the $^{1}\mathrm{S}_0$--$^{3}\mathrm{P}_2$ magnetic quadrupole (M2) transition~\cite{yamaguchi10,onishchenko19,trautmann23,bohman23}.
The interrogation of these two distinct clock transitions in the same atomic species can be advantageous for the evaluation of systematic shifts of the clock frequency~\cite{bohman23,furst20}, the search for physics beyond the standard model~\cite{dzuba18}, and as a testbed for high precision atomic structure calculations~\cite{SafKozJoh09}.

Recently, clock transitions have found promising applications in quantum computing with atom arrays, where metastable clock states can serve as excellent qubits and present a gateway for high-fidelity two-qubit gates~\cite{madjarov20,okuno22,lis23}.
Single-qubit gates have been demonstrated on the $^{1}\mathrm{S}_0$--$^{3}\mathrm{P}_0$ transition in two-electron atoms and second-scale atomic coherence times have been observed~\cite{madjarov20,young20}.
Intriguingly, the combination of the metastable states $^{3}\mathrm{P}_0$ and $^{3}\mathrm{P}_2$ has been proposed for the implementation of a fast, high-fidelity qubit~\cite{pagano22}.
In combination with the ground state, the realization of an all-optical qutrit $^{1}\mathrm{S}_0$--$^{3}\mathrm{P}_0$--$^{3}\mathrm{P}_2$ is conceivable~\cite{trautmann23}.

Quantum simulation experiments with ultracold atoms also benefit from the high frequency resolution provided by narrow optical transitions.
Here, precision spectroscopy allows manipulating and probing quantum many-body systems at previously inaccessible energy scales~\cite{zhang14,kato16}.
Because the nonspherical $^{3}\mathrm{P}_2$ state posseses a nonzero \emph{electric} quadrupole moment~\cite{derevianko01,santra04}, it provides a new platform to study quantum gases with anisotropic, long-range quadrupole-quadrupole interactions~\cite{bhongale13,lahrz14}.

For applications using ultracold atoms in metrology and quantum technology it is indispensable to preserve the inherently identical nature of the atoms by confining them in state insensitive optical potentials~\cite{katori03}. For the $^{3}\mathrm{P}_2$ state, such ``magic'' trapping conditions can be achieved over an extensive range of trap wavelengths, spanning hundreds of nanometers by careful adjustment of the lattice polarization~\cite{trautmann23}. In particular, this tunability enables operating at trapping wavelengths where high power laser technology is available, thus significantly raising prospects for the scalability of neutral atom quantum technologies~\cite{park22}.

In this Letter, we realize the first coherent precision spectroscopy of the ultranarrow $^{1}\mathrm{S}_0$--$^{3}\mathrm{P}_2$ M2 transition in $^{88}\mathrm{Sr}$.
We demonstrate high-fidelity coherent excitation to the $^{3}\mathrm{P}_2$ state, as well as high-resolution Rabi spectroscopy.
The coherence properties of our system are systematically evaluated by performing Ramsey spectroscopy. Finally, we use the obtained spectroscopy results to determine the linewidth of the M2 transition at the \textmu Hz scale. We compare our result to ab-initio theoretical predictions, which have awaited experimental verification for more than two decades~\cite{derevianko01}.

\paragraph{Experimental sequence.}  Our work starts by loading $2 \times 10^{5}$ $^{88}\mathrm{Sr}$ atoms into a three-dimensional optical lattice. As shown in Fig.~\ref{fig:setup}(b), the vertical lattice is formed by a retroreflected 1064-nm laser beam and the two horizontal lattice axes are generated by 914-nm laser beams inside an enhancement cavity~\cite{park22}. The resulting three-dimensional optical lattice has a depth of $150\,E_\mathrm{rec}$ ($270\,E_\mathrm{rec}$) and trap frequency of \unit{65}{kHz} (\unit{68}{kHz}) along the horizontal (vertical) axes, respectively, where $E_\mathrm{rec} = h^2/(2m\lambda_{\mathrm{l}}^2)$ is the lattice photon recoil energy for an atom of mass $m$ at the corresponding lattice wavelength $\lambda_{\mathrm{l}}$ and $h$ denotes Planck's constant. The average occupation per lattice site is below unity and tunneling is negligible on experimental time scales such that strong inelastic collisions between atoms in the $^{3}\mathrm{P}_2$ state can be avoided~\cite{traverso09,lisdat09}.

Despite the large disparity in wavelength of the different lattice axes, we achieve magic trapping conditions by matching the polarizability of the excited state $|e\rangle = |^{3}\mathrm{P}_2, m_J = 0\rangle$ to the ground state $|g\rangle = |^{1}\mathrm{S}_0\rangle$.
While the polarizability of $|g\rangle$ is solely determined by $\lambda_{\mathrm{l}}$, the polarizability of the nonspherical state $|e\rangle$ can be tuned via the angle $\beta$ between linear lattice polarisation and quantization axis according to~\cite{lekien13,heinz20}
\begin{equation}
	\label{eq:polarizability}
  \begin{aligned}
    \alpha(\lambda_{\mathrm{l}}, \beta)& =  \alpha_{\mathrm{s}}(\lambda_{\mathrm{l}}) - \alpha_{\mathrm{t}}(\lambda_{\mathrm{l}}) \dfrac{3 \cos^{2}\beta -1 }{2},\\
      \end{aligned}
\end{equation}
where $\alpha_{\mathrm{s}}$ and $\alpha_{\mathrm{t}}$ denote scalar and tensor polarizabilites, respectively. In our experiment, the quantization axis is defined by a \unit{28}{G} bias magnetic field $\mathbf{B}$ angled at the measured ``magic angle'' $\beta_{\mathrm{m,1064}} = 16(2)\degree$ with respect to the polarization of the 1064-nm vertical lattice~\cite{trautmann23}.
For the 914-nm lattices we find magic trapping conditions close to the theoretically predicted angle $\beta_{\mathrm{m,914}} = 52\degree$ (see Supplemental Material~\cite{supplemental}).

\begin{figure}
	\centering
	\includegraphics{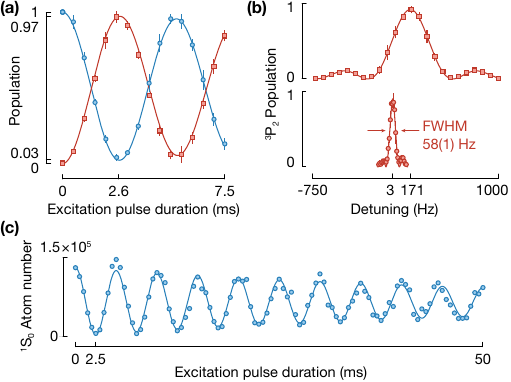}
	\caption{Rabi spectoscopy. (a) Rabi oscillations of the $|g\rangle$ (blue circles) and $|e\rangle$ (red squares) populations for varying excitation times. Solid lines are fits to an exponentially damped sinusoidal oscillation. From the amplitudes of the fits we extract a maximum excitation fraction of $97(1)\%$. (b) Spectra obtained for $\pi$ pulse durations of \unit{2.6}{ms} (top) and \unit{15}{ms} (bottom). Solid lines show fits to a Rabi lineshape with FWHM of \unit{370(2)}{Hz} and \unit{58(1)}{Hz}, respectively. We observe a probe-light-induced Stark shift of the line center and set the zero of the frequency axis to the unshifted transition frequency. (c) Long-term evolution of Rabi oscillations displaying motional dephasing due to the finite temperature of the atomic sample. The solid line shows a fit to an exponentially damped oscillation yielding a $1/e$ decay time of \unit{57(5)}{ms}.} \label{fig:rabi}
\end{figure}

\begin{figure*}[t]
  \centering
  \includegraphics{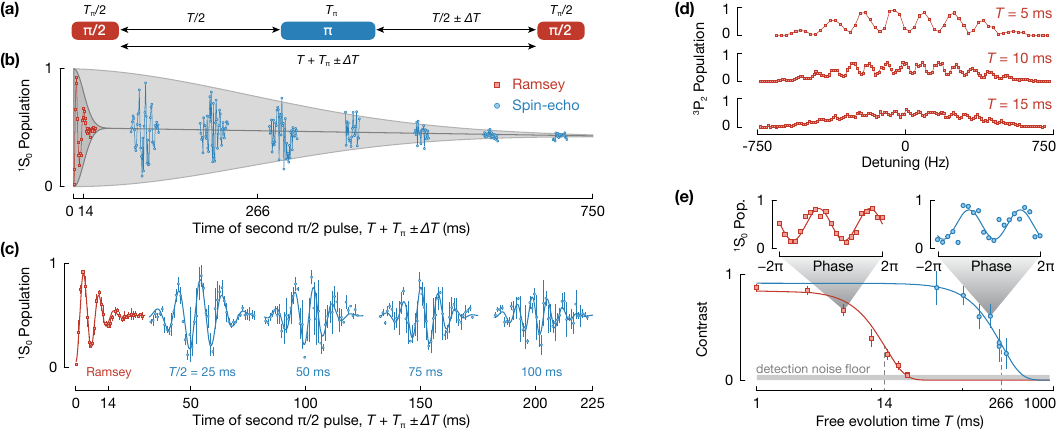}
  \caption{Ramsey spectroscopy. (a) Illustration of pulse sequences employed for Ramsey and spin-echo experiments. (b) Ramsey (red squares) and spin-echo (blue circles) signals from single experimental runs. The dark (light) gray shaded regions represent a Gaussian decay of the Ramsey (spin-echo) contrast combined with an exponential decay to account for finite atom lifetime in the lattice. (c) Ramsey and spin-echo signals averaged over three experimental runs. The solid lines represent fits to a sinusoidal oscillation with Gaussian envelope to guide the eye. (d) Ramsey fringes for free evolution times $T$ of \unit{5}{ms}, \unit{10}{ms}, and \unit{15}{ms} (from top to bottom). (e) Contrast decay of the Ramsey (red squares) and spin-echo (blue circles) signals for different free evolution times $T$. The gray-shaded area denotes the detection noise floor of our system set by atom number fluctuations and the insets show Ramsey and spin-echo phase scans. The solid lines represent Gaussian fits to the contrast decay, with $1/e$ decay times of \unit{14(1)}{ms} and \unit{266(36)}{ms} for Ramsey and spin-echo sequences, respectively, leading to the gray-shaded areas in panel (b).}
  \label{fig:ramsey}
\end{figure*}

After transferring the atoms to the lattice, we apply resolved sideband cooling on the $^{1}\mathrm{S}_0$--$^{3}\mathrm{P}_1$ transition.
The resulting temperature of \unit{2.3}{\mu K} (\unit{2.1}{\mu K}) corresponds to an average phonon occupation of $\bar{n} = 0.35$ ($0.27$) and a ground state fraction of $74\%$ (79\%) along the probe (vertical) axis.

To interrogate the $^{1}\mathrm{S}_0$--$^{3}\mathrm{P}_2$ M2 transition, we apply a 671 nm probe beam propagating with wave-vector $\mathbf{k}$ along the horizontal \textit{x}-axis. We use probe powers of up to \unit{50}{mW} and a $1/e^2$ beam waist radius of \unit{500}{\mu m}, much larger than the sample size of about \unit{50}{\mu m}. The linewidth of the frequency-stabilized probe laser is on the Hertz level~\cite{supplemental}. Its polarization $\bm{\epsilon}$ is chosen to be linear and is oriented orthogonal to the plane spanned by $\mathbf{k}$ and $\mathbf{B}$. We note that this choice of polarization excludes excitation of the $^{1}\mathrm{S}_0$--$^{3}\mathrm{P}_2$ transition as an electric-dipole (E1) transition induced via magnetic field admixing~\cite{taichenachev06}.

Following the spectroscopy sequence, we read out the number of $|g\rangle$ atoms through absorption imaging on the $^{1}\mathrm{S}_0$--$^{1}\mathrm{P}_1$ transition.
Alternatively, after removal of $|g\rangle$ atoms, the number of $|e\rangle$ atoms is measured by repumping atoms to $|g\rangle$ via the $^{3}\mathrm{S}_1$ state and taking an absorption image [see Fig.~\ref{fig:setup}(a)].
Populations are obtained via normalization to interleaved reference measurements of the total atom numer~\cite{supplemental}.

\paragraph{Rabi spectroscopy.}  The realization of a state-insensitive trap allows us to perform precision spectroscopy on the ultranarrow M2 transition. We coherently excite atoms to $|e\rangle$ by performing Rabi spectroscopy. Upon applying square probe pulses of varying pulse duration, we observe coherent oscillations of the atomic state populations, as shown in Fig.~\ref{fig:rabi}(a) for a Rabi frequency of $\Omega = 2\pi\times\unit{189(1)}{Hz}$. For a \unit{2.6}{ms}-long resonant $\pi$ pulse we observe an excitation fraction of $97(1)\%$.

In Fig.~\ref{fig:rabi}(b), we show Rabi spectra obtained for two different probe beam intensities. We adjust the probe pulse duration to perform excitation with a $\pi$ pulse when on resonance. For pulse durations of \unit{2.6}{ms} and \unit{15}{ms}, we observe spectra with \unit{370(2)}{Hz} and \unit{58(1)}{Hz} full-width-at-half-maximum (FWHM), respectively.
The minimum achievable linewidth is limited by residual lattice light shifts and eventually by the finite linewidth of the probe laser.
In addition to the difference in linewidths, we also observe a shift in the resonance frequency for different probe beam intensities.
This probe-light-induced differential ac Stark shift $\Delta_{\mathrm{ac}}$ arises due to off-resonant coupling of the probe light to additional states~\cite{taichenachev06}.

The long-term evolution of Rabi oscillations at a fitted Rabi frequency of $\Omega = 2\pi\times\unit{200(1)}{Hz}$ is shown in Fig.~\ref{fig:rabi}(c). We observe damping with a $1/e$ timescale of \unit{57(5)}{ms}, limited predominantly by motional dephasing due to the finite temperature and confinement of the atomic sample along the interrogation axis. This dephasing arises because after sideband cooling several vibrational states remain populated and the carrier Rabi frequency depends on the vibrational state~\cite{blatt09,kaufman12}. Other effects that contribute to the dephasing are residual lattice light shifts, intensity fluctuations of the probe beam, and spatial inhomogeneities (see Supplemental Material~\cite{supplemental}).

\paragraph{Ramsey spectroscopy.}  To characterize the atom--atom coherence of the prepared states we perform Ramsey spectroscopy and optionally add a spin-echo pulse, as sketched in Fig.~\ref{fig:ramsey}(a).
For Ramsey spectroscopy, we apply a $\pi / 2$ pulse of duration $T_\pi / 2 = \unit{1.3}{ms}$ to prepare a coherent superposition state $(|g\rangle + |e\rangle)/\sqrt{2}$, which evolves to $(|g\rangle + e^{-i\phi} |e\rangle)/\sqrt{2}$ by acquiring a relative phase shift $\phi$ during the free evolution time $T$.
Finally, we map this phase difference into a population difference by applying another $\pi / 2$ pulse to obtain the state  $(i\sin\phi|g\rangle + \cos\phi |e\rangle)/\sqrt{2}$~\cite{marti18}.
In our experiment, the acquired phase is proportional to the probe-light-induced Stark shift $\Delta_\mathrm{ac}$.
For the probe intensity used in Fig.~\ref{fig:ramsey}, the resulting Ramsey signal oscillates at about \unit{140}{Hz}.
The oscillations exhibit a Gaussian decay indicating an inhomogeneous distribution of detunings, leading to the loss of atom--atom coherence. We also observe this contrast decay for Ramsey fringes as a function of detuning for increasing $T$, as shown in Fig.~\ref{fig:ramsey}(d).

To quantify these results, we measure the decay of the Ramsey contrast when scanning the phase of the final $\pi / 2$ pulse for varying $T$ in Fig.~\ref{fig:ramsey}(e). We extract an inhomogeneous dephasing time $T_{2}^* = 14(1)~\mathrm{ms}$, indicating a Gaussian distribution of detunings with a standard deviation of $\sqrt{2}/T_{2}^* = 2\pi\times16(1)~\mathrm{Hz}$~\cite{madjarov20}. This inhomogenous distribution can be mainly attributed to residual light shifts of the optical lattice and sets a limit for the most narrow observable spectra~\cite{supplemental}.

Next, we demonstrate that dephasing of the atom--atom coherence due to lattice light shifts can be reversed to a large extent through the application of a spin-echo sequence~\cite{kuhr05}. To this end, we extend the spectroscopy sequence by an additional $\pi$ pulse of duration $T_\pi$ after a free evolution time $T/2$ [see Fig.~\ref{fig:ramsey}(a)].
In Fig.~\ref{fig:ramsey}(b) and (c) we observe rephasing of the atomic spins in the form of a spin echo at time $T$ by scanning the timing of the final $\pi/2$ pulse.
For evolution times within the coherence time of our clock laser, we observe spin-echo signals that feature coherent oscillations.
Even after atom--laser coherence is lost, we still observe variance in the spin-echo signal for longer evolution times. This confirms the presence of atom--atom coherence, which is now probed with a random phase of the clock laser~\cite{young20}. From the Gaussian decay of phase contrast measurements shown in Fig.~\ref{fig:ramsey}(e) we infer a homogeneous dephasing time $T_{2}' = \unit{266(36)}{ms}$~\cite{kuhr05}, presumably limited by fluctuations of the quantization axis angle and of the lattice polarizations~\cite{supplemental}. For even longer evolution times, we start to observe the finite excited-state trap lifetime with a $1/e$ time scale of \unit{5}{s} due to trap-induced Raman scattering~\cite{dorscher18} and the onset of tunneling leading to collisional loss of atoms in $|e\rangle$~\cite{traverso09,lisdat09}.

\begin{figure}[t]
	\centering
	\includegraphics{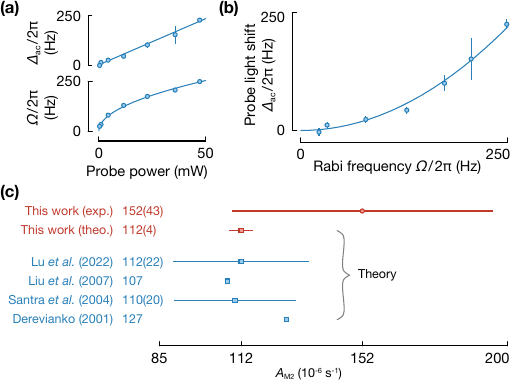}
	\caption{Evaluation of M2 transition rate. (a) Measured probe light shifts $\Delta_{\mathrm{ac}}$ (top) and Rabi frequencies $\Omega$ (bottom) for different probe beam powers. Solid lines are fits denoting a linear and square root scaling, respectively. (b) From a nonlinear fit of the form $\Delta_{\mathrm{ac}} = \xi^{-1}\Omega^2$ with reduced $\chi^2$ of 1.1 we extract the relative excitation strength $\xi = 2\pi\times\unit{283(12)}{Hz}$. (c) Comparison of experimental (circle) and theoretical (squares) values of the decay rate $\mathrm{A_{M2}}$~\cite{derevianko01,santra04,liu07,lu22}.}
	\label{fig:matrixelement}
\end{figure}

\paragraph{Transition linewidth.}  Finally, we use the obtained spectroscopic results to determine the linewidth of the M2 transition by extracting the transition rate $A_\mathrm{M2}$ for emission of a magnetic quadrupole photon. Simultaneous measurement of the Rabi frequency $\Omega$ along with the probe light shift $\Delta_{\mathrm{ac}}$ allows for an accurate determination of the transition strength insensitive to the probe beam intensity.
Following Lange~\textit{et al.}~\cite{lange21}, we define the probe-intensity-independent quantity $\xi = \Omega^{2}/\Delta_{\mathrm{ac}}$, called the relative excitation strength. In Fig.~\ref{fig:matrixelement}, we plot $\Omega$ and $\Delta_{\mathrm{ac}}$ for different intensity settings of the probe beam.
From a nonlinear fit to this data we extract a relative excitation strength $\xi = 2\pi\times\unit{283(12)}{Hz}$~\cite{supplemental}.

We combine $\xi$ with the differential polarizability $\Delta\alpha_{\mathrm{eg}}(\nu_0)$ at the transition frequency $\nu_0$ and determine the transition rate~\cite{lange21,supplemental} as
\begin{equation}
\label{eq:decayrate}
A_{\mathrm{M2}} = \dfrac{8\pi\nu_0^3}{5\epsilon_0 c^3}\Delta\alpha_{\mathrm{eg}}(\nu_0)\xi\dfrac{1}{6\cos^2\theta\sin^2\theta},
\end{equation}
where $\epsilon_0$ is the vacuum permittivity and $c$ is the speed of light.
The last term models the geometry of the probe beam propagating at an angle $\theta$ relative to the quantization axis with linear polarization perpendicular to the plane spanned by $\mathbf{k}$ and $\mathbf{B}$~\cite{schulz20}.
For the differential polarizability at the transition frequency, we use the theory value $\Delta\alpha_{\mathrm{eg}}(\nu_0) = \unit{290(7)}{a.u.}$~\cite{supplemental}.
Using these results with our experimental angle $\theta = 14(2)\degree$~\cite{supplemental} and the absolute transition frequency $\nu_0$~\cite{trautmann23}, the decay rate of the M2 transition in $^{88}\mathrm{Sr}$ is
\begin{equation}
\label{eq:decayratevalue}
A_{\mathrm{M2}} = \unit{152(43) \times 10^{-6}}{s^{-1}},
\end{equation}
corresponding to a transition linewidth $A_\mathrm{M2}/(2\pi) = \unit{24(7)}{\mu Hz}$ and a lifetime for emission of magnetic quadrupole radiation of $A_\mathrm{M2}^{-1} = \unit{110(31)}{min}$. The uncertainty is dominated by uncertainties in the angle $\theta$~\cite{supplemental}.

We stress that to obtain the total lifetime of the $^{3}\mathrm{P}_2$ state, other relevant decay channels have to be considered. In particular, magnetic dipole (M1) decay to $^{3}\mathrm{P}_1$ and blackbody-radiation-induced quenching to $4d$ $^{3}\mathrm{D}_J$ states are expexted to contribute significantly at room temperature~\cite{derevianko01,yasuda04}.

In Fig.~\ref{fig:matrixelement}(c), we compare our experimentally measured transition rate to a new ab-initio calculation and to theory predictions from the literature~\cite{derevianko01,santra04,liu07,lu22}. To calculate $A_{\mathrm{M2}}$ we use a hybrid approach combining configuration interaction (CI) and coupled cluster (CC) methods~\cite{SafKozJoh09}. The CI computation  includes an interaction between the two valence electrons of strontium while the coupled cluster method accounts for core--core and core--valence correlations. The wave functions $\Phi_n$ and energy levels $E_n$ of the valence electrons are found by solving the multiparticle relativistic equation $H_{\rm eff}(E_n) \Phi_n = E_n \Phi_n$~\cite{DzuFlaKoz96}, where the effective Hamiltonian is defined as $H_{\rm eff}(E) = H_{\rm FC} + \Sigma(E)$ with the Hamiltonian $H_{\rm FC}$ in the frozen-core approximation. The energy-dependent operator $\Sigma(E)$ accounts for virtual excitations of the core electrons. To quantify uncertainties, which arise from incomplete inclusion of the core, we carry out several computations, which include the core, i.e. build  the $\Sigma(E)$ operator, in different approximations (see Supplemental Material~\cite{supplemental}). As a result, we estimate the absolute value of the M2 $\langle ^1\mathrm{S}_0\|\mathrm{M2}\|^3\mathrm{P}_2\rangle $ matrix element to \unit{22.6(4)}{\mu_B} with 2\% total uncertainty and the M2 transition rate to \unit{112(4) \times 10^{-6}}{s^{-1}} with 4\% uncertainty. Here, $\mu_B$ denotes the Bohr magneton.

\paragraph{Summary and outlook.}  We have demonstrated that the $^{1}\mathrm{S}_0$--$^{3}\mathrm{P}_2$ M2 transition can be coherently controlled with high frequency resolution.
Excitation to the $^{3}\mathrm{P}_2$ state is possible with high fidelity and we have observed hundreds of milliseconds long coherence times, despite the higher sensitivity to environmental perturbations of the $^{3}\mathrm{P}_2$ state compared to the well-established $^{3}\mathrm{P}_0$ state.
These results enable applications of the $^{3}\mathrm{P}_2$ state in quantum information processing~\cite{pagano22}.
Further improvements in pulse fidelities and coherence times are expected by improved cooling or by employing robust qubit rotation and dynamical decoupling schemes~\cite{bluvstein22}.
Reaching lower temperatures will also enable operation at lower trap depths where lattice light shifts are even further reduced.
With the obtained advances in spectral resolution, many energy scales of quantum many-body systems, such as tunneling dynamics or contact interactions, can now be probed on the M2 transition.
The study of long-range quadrupolar interactions between ultracold atoms in the $^{3}\mathrm{P}_2$ state, which give rise to frequency shifts on the Hertz level, is now within reach~\cite{bhongale13,lahrz14}.
Our results establish an additional clock transition in strontium, demonstrate the suitability of the $^{3}\mathrm{P}_2$ state for a qubit, and pave the way for a unique new quantum simulation platform based on nonspherical atoms.

\begin{acknowledgments}
  We thank A.\,Derevianko for stimulating discussions, S.\,Snigirev and Y.\,Yang for support with experimental software and hardware, and A.\,Schindewolf and J.\,Geiger for careful reading of the manuscript. The Munich team acknowledges funding by the Munich Quantum Valley initiative as part of the High-Tech Agenda Plus of the Bavarian State Government, by the BMBF through the program ``Quantum technologies -- from basic research to market'' (Grant No. 13N16357), and funding under the Horizon Europe program HORIZON-CL4-2022-QUANTUM-02-SGA via the project 101113690 (PASQuanS2.1).
The Delaware team's research was supported in part by the USA Office of Naval Research (Grant Number N00014-20-1-2513), the European Research Council (ERC) under the European Union's Horizon 2020 research and innovation program (Grant Number 856415), and the USA NSF QLCI Award OMA – 2016244 and through the use of University of Delaware HPC Caviness and DARWIN computing systems. V.\,K. thanks the Hector Fellow Academy for support.
\end{acknowledgments}

\appendix


%

\end{document}


\title{Supplemental Material: \\
  Coherent excitation of a \textmu Hz scale optical magnetic quadrupole transition}

\author{V. Kl{\"u}sener}
\author{S. Pucher}
\affiliation{
  Max-Planck-Institut f{\"u}r Quantenoptik,
  85748 Garching, Germany}
\affiliation{
  Munich Center for Quantum Science and Technology,
  80799 M{\"u}nchen, Germany}
\author{D. Yankelev}
\thanks{Current address: Rafael Ltd, Haifa 3102102, Israel}
\affiliation{
  Max-Planck-Institut f{\"u}r Quantenoptik,
  85748 Garching, Germany}
\affiliation{
  Munich Center for Quantum Science and Technology,
  80799 M{\"u}nchen, Germany}
\author{J. Trautmann}
\author{F. Spriestersbach}
\affiliation{
  Max-Planck-Institut f{\"u}r Quantenoptik,
  85748 Garching, Germany}
\affiliation{
  Munich Center for Quantum Science and Technology,
  80799 M{\"u}nchen, Germany}
\author{D. Filin}
\author{S. G. Porsev}
\author{M. S. Safronova}
\affiliation{
  Department of Physics and Astronomy,
  University of Delaware,
  Newark, Delaware 19716, USA}
\author{I. Bloch}
\affiliation{
  Max-Planck-Institut f{\"u}r Quantenoptik,
  85748 Garching, Germany}
\affiliation{
  Fakult{\"a}t f{\"u}r Physik,
  Ludwig-Maximilians-Universit{\"a}t M{\"u}nchen,
  80799 M{\"u}nchen, Germany}
\affiliation{
  Munich Center for Quantum Science and Technology,
  80799 M{\"u}nchen, Germany}
\author{S. Blatt}
\email{sebastian.blatt@mpq.mpg.de}
\affiliation{
  Max-Planck-Institut f{\"u}r Quantenoptik,
  85748 Garching, Germany}
\affiliation{
  Fakult{\"a}t f{\"u}r Physik,
  Ludwig-Maximilians-Universit{\"a}t M{\"u}nchen,
  80799 M{\"u}nchen, Germany}
\affiliation{
  Munich Center for Quantum Science and Technology,
  80799 M{\"u}nchen, Germany}

\date{\today}

\maketitle

\section{Experimental procedure}
\label{sec:procedure}

The experimental setup has been described previously~\cite{park22,trautmann23}. Notable improvements include combining our cavity-enhanced horizontal lattices~\cite{park22} with a retroreflected vertical lattice~\cite{trautmann23} to form a three-dimensional optical lattice, as well as the addition of a narrow-linewidth spectroscopy laser system. In brief, the experimental sequence consists of optical transport of the atoms from the magneto-optical trap (MOT) region to a second vacuum chamber. There, the atoms are adiabatically transferred to the linearly polarized vertical optical lattice at a wavelength of \unit{1064}{nm} and a first stage of sideband cooling along the axial direction is applied. Subsequently, the atoms are loaded into a 3D lattice by switching on the linearly polarized horizontal lattices at a wavelength of \unit{914}{nm}, followed by sideband cooling along all lattice axes. The horizontal lattices are generated inside an enhancement cavity with a finesse of about 5000. To precisely adjust the lattice polarizations, we use wave plates in motorized rotation mounts. Due to the limited repeatability of the rotation angles (bidirectional repeatability $0.05\degree$) we choose to fix the lattice polarizations at the magic angles for the spectroscopy. Residual differential light shifts during sideband cooling stages limit the achievable ground state fraction to about $74\%$. We are actively exploring rotation mounts with an order of magnitude improved repeatability, which will enable precise dynamic polarization control and allow reaching ground state fractions exceeding $90\%$. Reaching lower temperatures will directly improve coherence times and enable operating at lower trap depths, where lattice light shifts are reduced.

\section{Data analysis}
\label{sec:analysis}

Unless stated otherwise, all errors and uncertainties denote one standard deviation confidence intervals. To correct for slow drifts in atom number, the Rabi spectroscopy measurements are interleaved by reference measurements without the application of a probe pulse. Interpolation of these reference measurements is used to correct the normalization of the raw atom number data~\cite{trautmann23}.

\section{Spectroscopy laser system}
\label{sec:laser}

The laser system for spectroscopy of the $^{1}\mathrm{S}_0$-$^{3}\mathrm{P}_2$ transition is based on a homebuilt linear external cavity diode laser (ECDL) operating at a wavelength of \unit{671}{nm}. Frequency stabilization is performed via locking to an ultrastable reference cavity with a finesse of 180,000 using the Pound-Drever-Hall technique. The stabilized light is used to seed an injection locked diode from which we retrieve up to \unit{50}{mW} optical power at the position of the atoms. We employ a fiber noise cancellation system for the optical fiber between seed laser and injection locked diode to suppress phase noise in the low frequency regime. The resulting linewidth of the spectroscopy laser is evaluated by observing a heterodyne beat note with an independent laser source at the same wavelength, stabilized to a sub-Hz linewidth optical frequency comb. With fiber noise cancellation we observe beat notes with full width at half maximum (FWHM) of about \unit{1.5}{Hz}. The laser displays longterm linear frequency drifts corresponding to a drift rate of \unit{60}{mHz/s}. Due to its short-term nonlinear frequency drifts we interleave precision resonance frequency measurements, such as for the probe light shift, with reference spectra at fixed probe power to track and remove laser drifts.
We assume that the frequency of the spectroscopy laser drifts linearly between successive reference spectra and we conservatively estimate the error bar on the drift-compensated resonance frequency as the peak-to-peak variation between successive reference spectra.

\section{Polarizability}
\label{sec:polarizability}

The dynamic dipole polarizability $\alpha$ of an atomic state can be decomposed into a scalar polarizability $\alpha_{\mathrm{s}}$, a vector polarizability $\alpha_{\mathrm{v}}$, and a tensor polarizability $\alpha_{\mathrm{t}}$,
\begin{align}
\begin{split}
\alpha(\lambda,\gamma,\beta) = &\alpha_{\mathrm{s}}(\lambda) + \alpha_{\mathrm{v}}(\lambda) \sin(2 \gamma) \dfrac{m_{J}}{2 J}\\
&+ \alpha_{\mathrm{t}}(\lambda) \dfrac{3 \cos^{2}(\beta) -1 }{2} \dfrac{3m_{J}^{2} - J (J+1)}{J (2J -1)},
\end{split}
\end{align}
where $\lambda$ is the wavelength of interest, $\gamma$ is the ellipticity angle of the polarization, and $\cos\beta$ is the projection of the polarization vector onto the quantization axis~\cite{lekien13,heinz20}. Because the ground state $^{1}\mathrm{S}_0$ posseses no angular momentum ($J=0$), its polarizability is solely determined by the scalar polarizability $\alpha_{\mathrm{s}}$, which only depends on the trap wavelength $\lambda$. In contrast, for the metastable $^{3}\mathrm{P}_2$ state ($J=2$), a tensor contribution $\alpha_{\mathrm{t}}$ to the polarizability exists. In this work, we perform spectroscopy of the $^{3}\mathrm{P}_2$ $( m_J = 0)$ state using linear trap polarizations ($\gamma=0$), such that the vector contribution vanishes. The matrix elements required for calculating the polarizability of the $^{1}\mathrm{S}_0$ and $^{3}\mathrm{P}_2$ states have been presented previously~\cite{trautmann23}. In Table~\ref{tab:polarizability}, we summarize the resulting polarizability values at the trap wavelengths of \unit{914}{nm} and \unit{1064}{nm} used in this work. In addition, we present the polarizability at the probe wavelength of \unit{671.2}{nm}, which is required to model the probe-induced light shift for the determination of the M2 transition rate.

\begin{table}[h]
	\centering
	\caption{Polarizabilities of the $5s^{2}$ $^{1}\mathrm{S}_0$ state and the $5s5p$ $^{3}\mathrm{P}_2$ state of $^{88}\mathrm{Sr}$ at various wavelengths $\lambda$. The values are obtained from the energies and transition matrix elements presented in Ref.~\cite{trautmann23}. The atomic unit of polarizability is defined as 1 a.u. = $4\pi\epsilon_{0}a_{0}^{3}$, where $\epsilon_0$ is the vacuum permittivity and $a_0$ is the Bohr radius.}
	\begin{tabularx}{\columnwidth}{%
      @{}
      >{\RaggedRight}X
      @{}
      >{\setlength\hsize{0.33\columnwidth}\RaggedLeft}X
      @{}
      >{\setlength\hsize{0.33\columnwidth}\RaggedLeft}X
      @{}}
		\hline \hline
		\\
		State & $\alpha_{\mathrm{s}}(\lambda)$ & $\alpha_{\mathrm{t}}(\lambda)$ \\
		& (a.u.) & (a.u.) \\
		\hline
		\\
		\unit{671.2}{nm} & & \\
		$5s^{2}$ $^{1}\mathrm{S}_0$ & 359.1(1.7) & 0 \\
		$5s5p$ $^{3}\mathrm{P}_2$ & -142(6) & 422(4) \\
		\\
		\unit{914}{nm} & & \\
		$5s^{2}$ $^{1}\mathrm{S}_0$ & 261.1(1.3) & 0 \\
		$5s5p$ $^{3}\mathrm{P}_2$ & 255(4) & -77.1(1.4) \\
		\\
		\unit{1064}{nm} & & \\
		$5s^{2}$ $^{1}\mathrm{S}_0$ & 240.5(1.2) & 0 \\
		$5s5p$ $^{3}\mathrm{P}_2$ & 197(4) & -48.2(1.2) \\
		\\

		\hline \hline
	\end{tabularx}

	\label{tab:polarizability}
\end{table}

\begin{figure}
  \centering
  \includegraphics{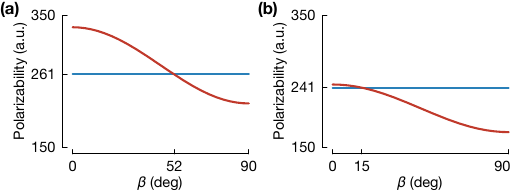}
  \caption{Theoretical polarizability of the $^{1}\mathrm{S}_0$ state (blue) and the $^{3}\mathrm{P}_2$ $( m_J = 0)$ state (red) as a function of angle $\beta$ between polarization vector and quantization axis for trap wavelengths of (a) \unit{914}{nm} and (b) \unit{1064}{nm}. We find theoretical ``magic angles'' at $\beta_{m,914} = 52\degree$ and $\beta_{m,1064} = 15\degree$, respectively.}
  \label{fig:polarizability}
\end{figure}

Figure~\ref{fig:polarizability} shows the theoretical polarizability as a function of the angle $\beta$ between polarization vector and quantization axis. We extract ``magic angles'' $\beta_{m,\lambda}$, where the polarizability of the $^{3}\mathrm{P}_2$ $(m_J = 0)$ state is matched to the ground state polarizability.
These angles are given by $\beta_{m,914} = 52\degree$ and $\beta_{m,1064} = 15\degree$ for trap wavelengths of \unit{914}{nm} and \unit{1064}{nm}, respectively.
At \unit{1064}{nm} the crossing of ground and excited state polarizabilities is shallow.
For this reason, we experimentally confirmed the magic angle in Ref.~\cite{trautmann23} to $16(2)\degree$.
At \unit{914}{nm}, we find a magic angle close to the predicted value, with a precision limited by the experimental uncertainties on the lattice light polarizations.
To theoretically estimate the sensitivity to deviations from the magic angle, we take the derivative of the polarizability curve at the magic angle and obtain sensitivities of \unit{-113}{a.u./rad} and \unit{-36}{a.u./rad} at \unit{914}{nm} and \unit{1064}{nm}, respectively.
We note that lower sensitivities can be obtained by operating at trap wavelengths where magic angles are in the vicinity of 0 or 90 degrees.

For the probe-induced Stark shift we evaluate the differential polarizability at the probe wavelength of \unit{671.2}{nm}. The contribution of the tensor polarizability $\alpha_{\mathrm{t}}$ depends on the angle $\beta$ between linear probe beam polarization and the quantization axis defined by the bias magnetic field. For our probe beam geometry we choose $\beta = 90\degree$, which minimizes the differential polarizability to \unit{290(7)}{a.u.} We note that this choice of polarisation simultaneously maximizes the transition strength to the  $^{3}\mathrm{P}_2$ $(m_J = 0)$ state for the given angle of the bias magnetic field~\cite{trautmann23}.

\section{Decoherence mechanisms}
\label{sec:decoherence}

\begin{figure}
  \centering
  \includegraphics{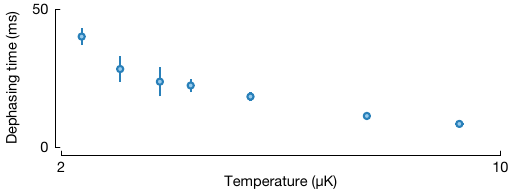}
  \caption{Measured $1/e$ dephasing time of Rabi oscillations at $\Omega=2\pi\times\unit{200}{Hz}$ for different temperatures of the atomic ensemble.}
  \label{fig:dephasing}
\end{figure}

We summarize different decoherence mechanisms present in our system and propose technical improvements to mitigate each effect. Unless stated otherwise, all simulations are performed for a Rabi frequency of $\Omega =2\pi\times\unit{200}{Hz}$.

\textit{Motional dephasing}.---After sideband cooling the atoms occupy several vibrational states $|n\rangle$ of the lattice. This leads to a dephasing of the observed Rabi oscillations because the carrier Rabi frequency depends on the vibrational level $|n\rangle$ according to
\begin{equation}
	\Omega(n) = \Omega_0 e^{-\eta^{2}/2} L_{n}(\eta^2),
\end{equation}
where $\Omega_0$ is the bare Rabi frequency, $\eta$ is the Lamb-Dicke factor, and $L_n$ denotes the Laguerre polynomial~\cite{kaufman12}. Assuming a thermal state at temperature $T = \unit{2.3}{\mu K}$ and using our experimental parameter $\eta = 0.28$ for the data presented in the main text, we derive a reduction of the $\pi$ pulse fidelity by \unit{0.7}{\%}. Figure~\ref{fig:dephasing} shows experimentally measured $1/e$ dephasing times $\tau_{\mathrm{Rabi}}$ for different temperatures $T$ of the atomic sample. We do not observe the theoretically expected beating and revival of the Rabi oscillations due to inhomogeneities in the trap frequency and the presence of additional dephasing mechanisms.

\textit{Lattice light shifts}.---Residual lattice light shifts due to deviations of the lattice polarizations from the magic angles lead to detuning inhomogeneities. We evaluate the magnitude of this effect from the contrast decay of the Ramsey signal, resulting in a Gaussian distribution of detunings with a standard deviation of \unit{16}{Hz}. Performing a Monte Carlo simulation of Rabi evolutions assuming this distribution of detunings, we deduce a reduction of the $\pi$ pulse fidelity by 0.6\%. Lattice light shifts are also the dominating decoherence effect that sets the inhomogeneous dephasing time $T_2^*$ of the atomic coherence. We demonstrated that the coherence time can be extended significantly through a spin-echo sequence. Further improvements in coherence time can be expected by employing more complex dynamical decoupling sequences~\cite{bluvstein22}, more precise lattice polarization control, or by operating at lower trap depths, which will be enabled by reaching lower sample temperatures.

\textit{Probe laser intensity noise}.---Low-frequency fluctuations of the probe beam intensity between experimental runs appear as dephasing due to a variation of the probe pulse area. In our setup, we find that the intensity pulse areas between different experimental runs are normally distributed with a fractional standard deviation of $\sigma_{\mathrm{RMS}}\approx0.7\%$. Approximating the noise in the Rabi frequency to also be normally distributed one derives a $1/e$ dephasing time of the Rabi oscillations of $\tau_{\mathrm{Rabi}}=2\sqrt{2}/(\Omega_{0}\sigma_{\mathrm{RMS}})$~\cite{madjarov20}. For $\Omega=2\pi\times\unit{200}{Hz}$ we therefore estimate $\tau_{\mathrm{Rabi}}\approx\unit{322}{ms}$ corresponding to a reduction of the $\pi$ pulse fidelity by 0.8\%. Higher gate fidelities can be expected with improved intensity stabilization techniques or by using robust single-qubit rotation schemes~\cite{bluvstein22}.

\textit{Probe beam inhomogeneity}.---The inhomogeneous Gaussian intensity profile of the probe beam across the atomic sample leads to a spatially inhomogeneous distribution of Rabi frequencies. In addition, it gives rise to a spatially dependent probe light shift. We use a probe beam with a waist of about \unit{500}{\mu m} and the atomic sample can be modelled as a Gaussian density distribution with a maximum extent of \unit{50}{\mu m} (one standard deviation) in the plane orthogonal to the probe beam. By performing a Monte Carlo simulation taking into account the spatial probe beam mode and the atomic density distribution, we estimate a reduction of the $\pi$ pulse fidelity by 0.1\%.

\textit{Magnetic field noise}.---The $^{1}\mathrm{S}_0$-$^{3}\mathrm{P}_2$ $(\Delta m = 0)$ transition is insensitive to first-order Zeeman shifts. However, due to the tensor light shift of the $^{3}\mathrm{P}_2$ state, a sensitivity to the angle of the bias magnetic field relative to the lattice polarizations remains. For our setup, we conservatively estimate the relative stability of three pairs of bias magnetic field coils to be on the order of $10^{-4}$. The resulting angle instability of tens of \textmu rad leads to a broadening of the $^{3}\mathrm{P}_2$ state on the order of a few Hertz for the lattice depths used in this work. This limits the homogeneous dephasing time $T_{2}'$ of the atomic coherence to below one second. Longer coherence times can be obtained by operating at lower trap depths, trapping wavelengths with a lower sensitivity to the magnetic field angle, employing dynamical decoupling sequences~\cite{bluvstein22}, and improved bias magnetic field stability, which is feasible with existing technology.

\begin{table}[h]
	\centering
	\caption{Simulated error budget for $\pi$ pulse fidelity at a Rabi frequency of $\Omega=2\pi\times\unit{200}{Hz}$.}
	\begin{tabularx}{\columnwidth}{%
      @{}
      >{\RaggedRight}X
      @{}
      >{\setlength\hsize{0.33\columnwidth}\RaggedLeft}X
      @{}}
		\hline \hline
		\\
		Error source & Contribution \\
		\hline
		\\
		Motional dephasing & \unit{0.7}{\%} \\
		Lattice light shift & \unit{0.6}{\%} \\
		Probe laser intensity noise & \unit{0.8}{\%} \\
		Probe beam inhomogeneity & \unit{0.1}{\%} \\
		\\
        \textbf{Total fidelity} & \textbf{\unit{97.8}{\%}} \\
		\hline \hline
	\end{tabularx}

	\label{tab:fidelity}
\end{table}

\section{Transition linewidth}
\label{sec:transitionlinewidth}

\textit{Theoretical model}.---In Ref.~\cite{lange21}, an expression for the transition matrix element $V_{\mathrm{eg}}$ of a general multipole transition in a many-electron atom has been derived. Atomic states are labelled by the total angular momenta $F_{g,e} = J_{g,e} + I$, their projections $m_{g,e}$, as well as additional quantum numbers $\gamma_{g,e}$. The matrix element for transitions between electronic  states with total electronic angular momenta $J_{g,e}$ and nuclear spin $I$ is then given by~\cite{lange21}
\begin{align}
\begin{split}
V_{\mathrm{eg}} = &\dfrac{\sqrt{\pi}ec}{2\pi\nu_0}i^{L}(-1)^{J_e+I+F_g+L}\sqrt{(2L+1)(2F_g+1)}\\
&\times
\begin{Bmatrix}
F_e & F_g & L\\
J_g & J_e & I \\
\end{Bmatrix}
\langle F_gm_gLm_e-m_g|F_em_e\rangle\\
&\times
\langle \gamma_eJ_e\|H_{\mathrm{ph}}(pL)\|\gamma_gJ_g\rangle \sum_{\lambda=\pm1}(i\lambda)^{p}d^{L}_{m_e-m_g,\lambda}(\theta),
\end{split}
\end{align}
where $e$ is the elementary charge, $c$ is the speed of light, and $\nu_0$ is the transition frequency. Here, $p=0$ ($p=1$) for a magnetic (electric) $2^L$-pole transition and $\langle \gamma_eJ_e||H_{\mathrm{ph}}(pL)||\gamma_gJ_g\rangle$ introduces a reduced matrix element by use of the Wigner-Eckart theorem. The last sum with the Wigner d function $d^{L}_{m_e-m_g,\lambda}(\theta)$ models the geometry of the probe beam with wave vector $\mathbf{k}$ propagating at an angle $\theta$ relative to the quantization axis of the atom. Here, linearly polarized probe light with polarization $\bm{\epsilon}$ in the plane spanned by $\mathbf{k}$ and the quantization axis is assumed~\cite{schulz20}.

For spectroscopy of bosonic isotopes of strontium, the theoretical model can be simplified significantly due to the lack of hyperfine strucuture ($I=0$) and because the ground state posseses no angular momentum ($J_g=0$). In our experiment we probe a magnetic quadrupole transition ($p = 0$, $L = 2$) to the $^{3}\mathrm{P}_2$ state ($J_e = m_e = 2$) with the probe beam polarization $\bm{\epsilon}$ orthogonal to the plane spanned by $\mathbf{k}$ and the quantization axis. Inserting these parameters and modifying the geometric factor according to our polarization configuration~\cite{schulz20}, the matrix element for the M2 transition is given by
\begin{align}
\label{eq:VegSr}
\begin{split}
V_{eg} = &-\dfrac{\sqrt{\pi}ec}{2\pi\nu_0}\\
&\times
\langle \gamma_eJ_e\|H_{\mathrm{ph}}(0,2)\|\gamma_gJ_g\rangle \sum_{\lambda=\pm1}\lambda d^{2}_{0,\lambda}(\theta).
\end{split}
\end{align}

The decay rate for emission of a photon with multipolarity ($pL$) is given by~\cite{lange21}
\begin{equation}
A = \dfrac{16\pi^2\alpha\nu_0}{2J_e+1}|\langle \gamma_eJ_e\|H_{\mathrm{ph}}(pL)\|\gamma_gJ_g\rangle|^2,
\end{equation}
with $\alpha$ the fine structure constant. We insert Eq.~\ref{eq:VegSr} and define the Rabi frequency as $\Omega = 2\pi E_0|V_{\mathrm{eg}}|/h$ and the probe light induced quadratic Stark shift as $\Delta_{\mathrm{ac}} = 2\pi E_0^2\Delta\alpha_{\mathrm{eg}}(\nu_0)/(2h)$, where $E_0$ is the electric field amplitude of the probe laser, $\Delta\alpha_{\mathrm{eg}}(\nu_0)$ is the differential polarizability at the transition frequency $\nu_0$, and $h$ denotes Planck's constant.

We explicitly evaluate the geometric factor encoded in the Wigner d function and arrive at the final expression for the M2 transition rate presented in the main text,
\begin{equation}
A_{\mathrm{M2}} = \dfrac{8\pi\nu_0^3}{5\epsilon_0 c^3}\Delta\alpha_{\mathrm{eg}}(\nu_0)\xi\dfrac{1}{6\cos^2\theta\sin^2\theta},
\end{equation}
with the relative excitation strength $\xi = \Omega^2/\Delta_{\mathrm{ac}}$.

\textit{Data analysis}.---To extract the relative excitation strength $\xi$ from our data, we perform a fit to the experimentally measured Rabi frequencies $\Omega$ and probe light induced line center shifts $\Delta_{\mathrm{ac}}$ obtained for different probe beam powers.
The error bars on $\Delta_\mathrm{ac}$ are dominated by the spectroscopy-laser-drift-compensation scheme based on interleaved reference spectra taken between every data point.
For this reason, the error bars on $\Delta_{\mathrm{ac}}$ dominate over the errorbars on $\Omega$ in our experiment, and we perform a weighted non-linear least squares fit of the form $\Delta_{\mathrm{ac}} = \xi^{-1}\Omega^2$ with $\xi$ as the only free fit parameter and weights determined by the errorbars on $\Delta_{\mathrm{ac}}$. The fit yields a reduced $\chi^2$ of 1.1.

To determine the angle $\theta$ experimentally, we perform free-space Rabi spectroscopy on the $^{1}\mathrm{S}_0$--$^{3}\mathrm{P}_1$ using the same probe beam path and quantization axis angle as for the M2 spectroscopy. By comparing the excitation strengths of $\sigma^+$, $\sigma^-$, and $\pi$ transitions, we are able to deduce the angle $\theta$. Note that $\theta \neq \beta_{m,1064}$ because the vertical lattice polarisation is not oriented exactly in the \textit{xz}-plane.

\textit{Theoretical calculation of matrix element and transition rate}.---To calculate the M2 matrix element and transition rate, we use a hybrid approach combining configuration interaction (CI) and coupled cluster (CC) methods~\cite{SafKozJoh09}, as described in the main text. To estimate the total uncertatinty in these values, we carry out several computations, which include the core in different approximation. First, we use second-order many-body perturbation theory (MBPT) over residual Coulomb interaction~\cite{DzuFlaKoz96}, which is the least accurate approximation. The result is listed in Table~\ref{tab:TH} as CI+MBPT. Next, we use the linearized coupled cluster single-double (LCCSD) method~\cite{SafKozJoh09}, which includes higher-order corrections (CI+LCCSD). Finally, we add further triple valence and core excitations to the coupled cluster method \cite{Th3}. This final result is listed as CI+CCSDT in Table~\ref{tab:TH}. The difference between these results is very small, only 1\%.

All of the results include a random-phase approximation (RPA) correction to the M2 operator, which is small, 0.13~$\mu_B$. All other small corrections to the M2 operator \cite{1998Corrections,1999Corrections1,1999Corrections2} were estimated to be below 1.5\%. As a result, we estimate the total uncertainty of the matrix element to be 2\%, giving 4\% uncertainty in the $^1\mathrm{S}_0$--$^3\mathrm{P}_2$ M2 transition rate. The transition rate is calculated using the final CI+CCSDT matrix element and the experimental transition energy \cite{NIST}.

\begin{table}[h]
\caption{Absolute value of the M2 $\langle ^1S_0\|\mathrm{M2}\|^3P_2\rangle $ matrix element in units of $\mu_B$ and corresponding M2 transition rate in \unit{10^{-6}}{s^{-1}}.}
\label{tab:TH}
\begin{ruledtabular}
\begin{tabular}{lcccc}
\multicolumn{1}{c}{}  &
\multicolumn{1}{c}{CI+MBPT} &
\multicolumn{1}{c}{CI+LCCSD}  &
\multicolumn{1}{c}{CI+CCSDT} &
\multicolumn{1}{c}{Final}
 \\
\hline
\\
$\langle ^1\mathrm{S}_0\|\mathrm{M2}\|^3\mathrm{P}_2 \rangle$& 22.342& 22.508& 22.622 &22.6(4)\\
A(M2)                             &109.3 &110.9& 112.0&112(4) \\
\\
\end{tabular}
\end{ruledtabular}
\end{table}


%